\documentclass[arxiv,onecolumn,floatfix,footinbib,showpacs,superscriptaddress]{revtex4-1}
\usepackage[english]{babel}
\usepackage[utf8]{inputenc}
\usepackage{graphicx}
\usepackage{fancyhdr}
\usepackage[active]{srcltx}
\usepackage{setspace}
\usepackage{float}
\usepackage{amsmath}
\usepackage{amsfonts}
\usepackage{natbib}
\usepackage{color}

\usepackage{hyperref}
\usepackage{cleveref}

\begin{document}
\title{Tunable magnetoresistance in spin-orbit coupled graphene junctions}

\author{Razieh Beiranvand}
\email{razieh.beiranvand@abru.ac.ir}
\affiliation{Physics group, Department of basic science, Ayatollah Boroujerdi university, Boroujerd, Iran}

\author{Hossein Hamzehpour}
\affiliation{Department of Physics, K.N. Toosi University of Technology, 
Tehran 15875-4416, Iran}
\affiliation{School of Physics, Institute for Research in Fundamental Sciences (IPM), Tehran 19395-5531, Iran}


\begin{abstract}

Using the Landauer-B\"utikker formalism, we study the graphene magneto-transport in the presence of Rashba spin-orbit interaction (RSOI). we show that the angle resolved transmission
probability in the proposed structures can be tuned
by the RSOI strength. The transmission spectrum show Klein tunneling in the parallel (P) magnetization configuration which can be blocked by the RSOI. This effect is also observable for the anti-parallel (AP) magnetization configuration in different incident angle.
The numerical results shows that the spin-polarized conductance strongly depends on the strength of the RSOI and can be generated by tuning the magnetic exchange field and RSOI strength. This spin-polarized conductance is a sensitive oscillatory function of the thickness of the RSO region. Because of the spin-flip effect, the junction shows a spin-valve effect with large and negative magnetoresistance (MR) and spin-magnetoresistance (SMR) in the presence of RSOI. When the RSOI is on, the frequency and amplitude of shot-noise and Fano factor's oscillations are also increased. These results can provide a way to extending the application of graphene-based junctions in spintronics.

\end{abstract}
\maketitle

\section{Introduction}

Spintronics is a combination of two fundamental properties of the electron, namely charge and spin. A spintronic device is strongly dependent of the degree of spin polarization in the current. Thus, one of the main purposes in the field of spintronics is how to obtain and manipulate the spin-polarized currents \cite{DasSarma2001,Zutic2004}.

The most well-known application in spintronics is the spin-valve effect, in which the magnetoresistance (MR) of junctions can be controlled by tuning the strength and orientation of the magnetic exchange field \cite{Kovalev2006,Brataas2006}. A memory-storage cell and a read head are two common examples of the giant magnetoresistance (GMR) applications which is extremely used in the industry. The GMR sandwich structures made of alternating ferromagnetic and nonmagnetic metal layers. The device resistance in such a structure can change from a small to a large value depending on the relative orientation of the magnetization in the magnetic layers. Recent activities in designing and manufacturing spintronic devices following two different scenarios. The first is finding new materials with larger spin polarization or making improvements in the existing devices for better spin filtering. The second, which it seems to be more important, focuses on finding novel and proper ways of both generation and utilization of spin-polarized currents. It seems that these two targets can be accessible using the graphene-based spintronic devices \cite{Hill2006,Brey2007,Zhai2008}. 

Graphene is a versatile material with interesting electronic \cite{Nov2004,Nov2005,Park2008,Nov2009,Masrour2017_1}, optical \cite{Bonaccorso2010,Avouris2010,Bao2012}, magnetic \cite{Zhang2005,Son2006,Nair2012,Pesin2012,Han2014,Zareyan2010,Jiang2015,Jiang2016,Wang2018,Masrour2017_2,Jabar2016,Jabar2017,Masrour2018}, and thermal characteristics \cite{Inglot2015,Dragoman2007,Zuev2009,Wei2009} that could utilize in many device applications. It is also an important material for spintronic aims since the concentration of carriers can be controlled by gate voltages (or applied chemical potential).
The spin diffusion lengths of graphene can reach to 0.1 mm because of weak spin-orbit coupling in its layer \cite{Kane1995}.

Although in pristine graphene, the strength of spin–orbit interaction (SOI) is weak \cite{Min2006,Hernando2006,Yao2007}, but there is a possibility of enhancing the intensity of SOI in graphene layer both theoretically and experimentally \cite{Hernando2006}. 
Symmetries can generate two kinds of SOIs in graphene: (i) The intrinsic SOI originates from the intra-atomic spin-orbit couplings, and (ii) The extrinsic Rashba spin-orbit interaction (RSOI) generates due to the structure inversion symmetry in the presence of a perpendicular external electric field or curvature of the graphene sheet.
The extrinsic term of SOI in graphene has been estimated to be in the rage of about 0.05–0.0011 meV that can be increased via curvature inducing in the sheet of graphene \cite{Hernando2006} . Indeed, in some experimental conditions, the extrinsic term of SOI can reach values up to 200 meV at room temperature \cite{Yu2008}. Since the extrinsic RSOI can be tuned by an applied electric field, graphene can extremely used in spintronics. The RSOI is the most promising tool for the spin control. To demonstrate the existence of proximity-induced RSOI in a sheet of graphene, the inverse Rashba-Edelstein effect was implemented. In this method, a DC voltage along graphene layer is measured by spin to charge current conversion. 
Motivated by recent developments in spintronics with the novel material, the spin-polarized transport in graphene-based junctions are currently attracting a great deal of attentions  \cite{Zhai2008,Brey2007,Rojas2009,Bai2010,Wu2011,Liu2012}.
Theoretical study have suggested that the spin-dependent Klein tunneling in graphene makes the magnetoresistance exhibits the periodic
oscillation features which may be beneficial for the GMR devices \cite{Liu2012}.
In this paper we show that the application of Rashba spin-orbit (RSO) coupling in the graphene-based junctions could lead to an interesting MR behavior, providing qualitative insight into their scattering procedures. In order to obtain a high enough MR ratio, an alternative approach is to use the RSO region as an interlayer between different ferromagnetic regions which provide a spin-mixing procedure influencing the motion of Dirac electrons in graphene locally.

The paper is organized as follows: In Sec. II the theoretical method based on Landauer-B\"utikker formalism  is shortly outlined. Sec. III devoted to presentation of modeled structure and physical analysis, whereas the subsequent section summarizes the main results and conclusions.

\section{Model and analysis}
\begin{figure}[h]
 \centering
   \begin{tabular}{c}
     \includegraphics[width=8cm, height=6cm]{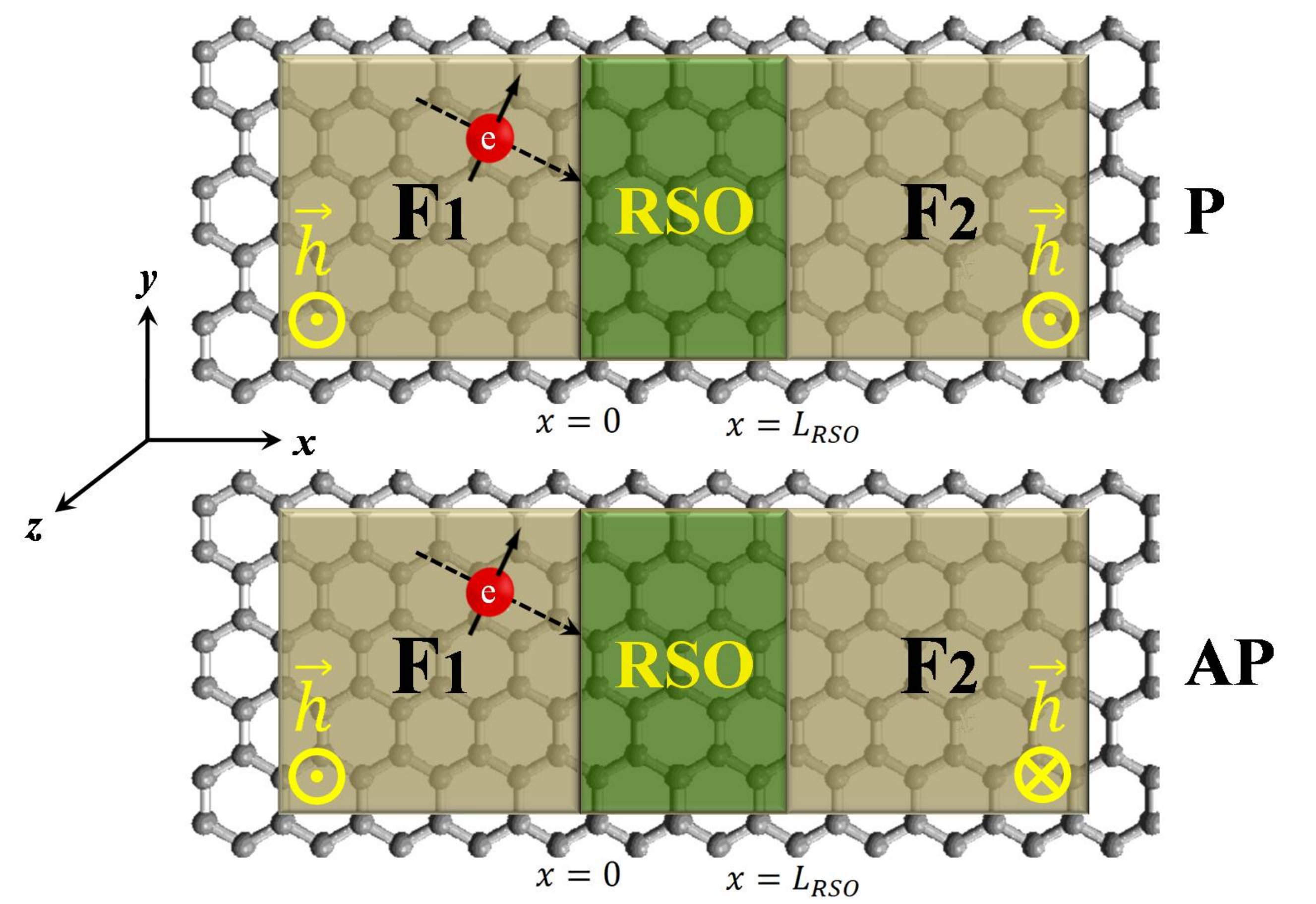}\\ 
   \end{tabular}
   \caption{A schematic illustration of the graphene-based F1-RSO-F2 junctions with parallel (P) and anti-parallel (AP) configuration of magnetic exchange field. The junctions are set in the $x-y$ plane and the uniform ferromagnetic and Rashba spin-orbit regions are supposed to be semi-infinite. A possible path of an incident spin-up electron at the interface is also shown in the F1 region. The F1-RSO-F2 boundaries are set at $x=0$ and $x=L_{RSO}$.}
   \label{Structures}
 \end{figure}

Let us, we consider a graphene-based F1-RSO-F2 junction made of two ferromagnetic (F) electrodes with a RSO region as a spin-mixing barrier as shown in Fig.{\ref{Structures}}. Since the ferromagnetism and RSOI can induce in graphene due to proximity \cite{Wang2015,Avsar2014}, such a structure can be realized experimentally by depositing F and RSO segments on the top of a graphene layer \cite{Wang2015}. 

The charge carriers in such systems can be described
by the following Dirac equation\cite{Beenakker2006},

\begin{equation}
       \left(\begin{array}{cc}
       H-\mu^i & 0\\
       0 & \mu^i-{\mathcal{T}}[H]{\mathcal{T}}^{-1}\\
       \end{array}\right)
       \left(\begin{array}{c}
        u \\
        v \\
       \end{array}\right)=\varepsilon
       \left(\begin{array}{c}
       u \\
       v \\
       \end{array}\right)\;,
       \label{Eq.D}
   \end{equation}
   
where $\mathcal{T}$ represents the time-reversal operator \cite{Beenakker2006}. The term $\mu^i$ refers to the chemical potential of each region that is easily tunable in graphene \cite{Nov2004}. In whole paper we set a constant value of chemical potential in all regions for simplicity.
$u$ and $v$ are the two-dimensional electron and hole spinors in one valley and $\varepsilon$ is the low-excitation energy of carriers. 

The Hamiltonian of the F1-RSO-F2 junction 
$H=\mathcal{H}_\text{F}+\mathcal{H}_\text{RSO}+\mathcal{H}_\text{D}$, consist of ferromagnetic part ($\mathcal{H}_\text{F}$) for $x\leq 0$ and $x\geq L_{RSO}$, Rashba spin-orbit part ($\mathcal{H}_\text{RSO}$) for $0\leq x\leq L_{RSO}$ and the two dimensional Dirac Hamiltonian in one valley $\mathcal{H}_D=s_0\otimes \hbar v_F \left(\sigma_x k_x+\sigma_y k_y\right)$ \cite{Beenakker2008}. 
The low-energy excitation in F segment is described by the Dirac-type equation, $\mathcal{H}_\text{F} = 
\left(s_z\otimes\sigma_0\right) h$ in which $h$ refers to the strength of the magnetic exchange field which added to the Dirac Hamiltonian via the Stoner approach. For simplicity, we assume that in both ferromagnetic regions, the magnetic exchange field is oriented in the $z$-direction, without loss of generality \cite{Halterman2013}. 
In the above equations, $k_x$ and $k_y$ are the components of wave vector in the $x$ and $y$ directions, respectively. $s_i$ and $\sigma_i$ are Pauli matrices acting on the real spin and pseudo-spin spaces related to the two triangular sub-lattices in the honeycomb lattice of graphene.
$s_0$ and $\sigma_0$ are $2\times 2$ unit matrices, and for simplicity we assume $\hbar v_F=1$. In case of graphene, because of the valley degeneracy, the final results are multiplied by 2.
In this case, we have two types of configurations namely parallel (P), in which $h_1$ and $h_2$ are in same directions with respect to the $z$-direction, and anti-parallel (AP), in which the directions of magnetization in the F1 and F2 are opposite. 

By diagonalizing the Dirac equation in the F region, eight eigenvalues are obtained,
  \begin{equation}
  \varepsilon_i=\pm \mu \pm\sqrt{(k^{\text{F}_i}_x)^2+(k_y)^2}\pm h_i \;,
  \end{equation} 
where $i=1,2$ is the index of F regions.

In the RSO region, the Hamiltonian $H$, in Eq. \ref{Eq.D} consists of two parts named $\mathcal{H}_D$ and $\mathcal{H}_\text{RSO}=-\lambda \left(s_y\otimes\sigma_x-s_x\otimes\sigma_y\right)$ in which $\lambda$ is the strength of RSOI. By diagonalizing the Hamiltonian of RSO region, we find following dispersion relation,
 \begin{equation}
 \begin{array}{cc}
  \varepsilon=\pm\mu+\zeta \sqrt{(k_x^\text{RSO})^2+(k_y)^2+
  \lambda^2}+ \eta \lambda,
  \label{EigenHSO}
 \end{array}
 \end{equation}
 
in which the band indices indicated by $\eta,\zeta=\pm 1 $.  The band structure in the presence of RSOI is gap-less with a splitting of magnitude $2\lambda$ between two sub-bands in contrast to the intrinsic spin-orbit couplings. The sub-band splitting due to the RSOI results in very interesting phenomena \cite{Beiranvand2016,Beiranvand2017,BeiranvandJAP2017}.

\begin{figure}[h]
 \centering
   \begin{tabular}{c}
     \includegraphics[width=10cm, height=6cm]{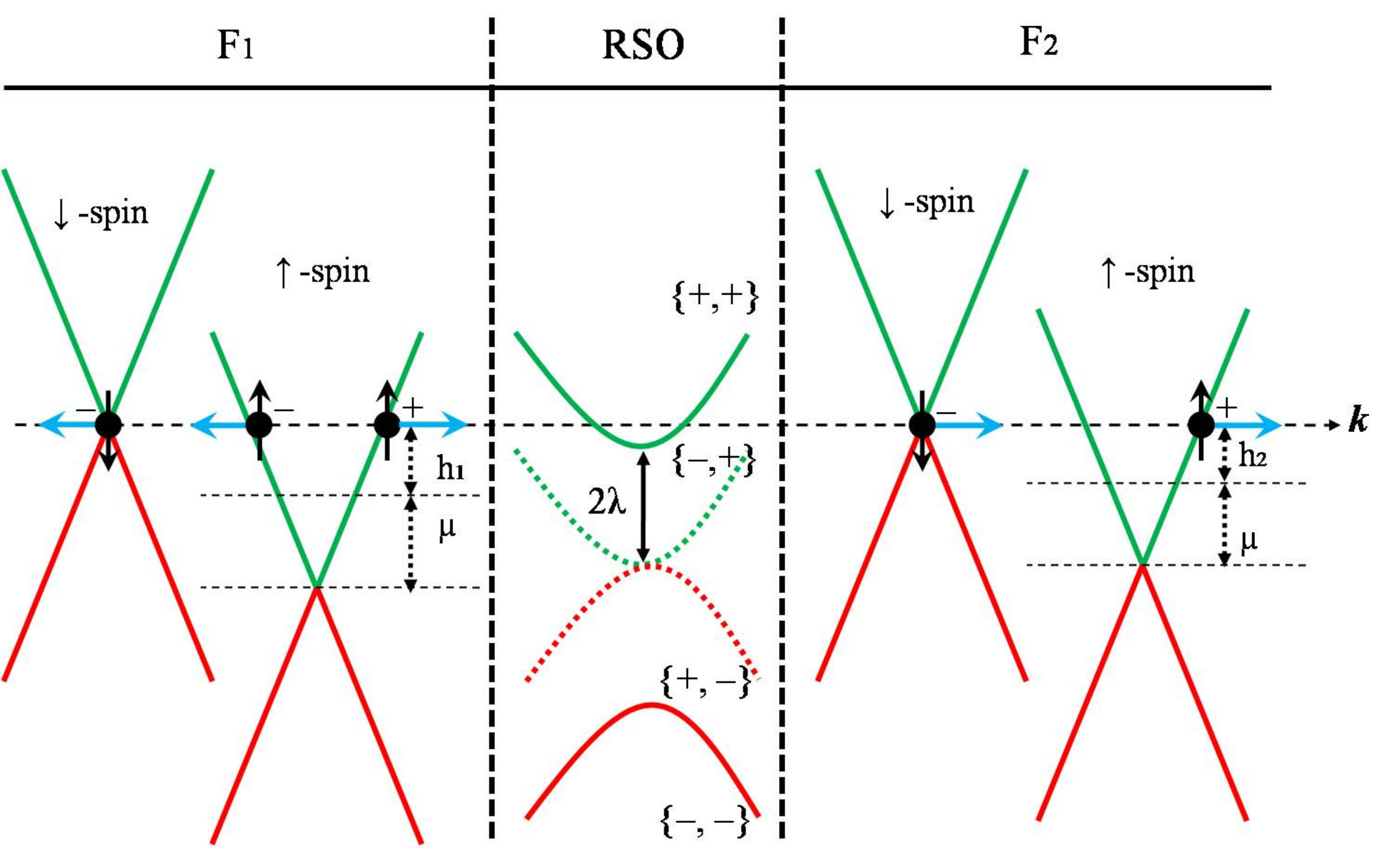}\\ 
   \end{tabular}
   \caption{Low energy band structure in each region. In the F regions, $\uparrow$-spin and $\downarrow$-spin electrons belong to different subbands. In the RSO region, there are two subbands splited by $2\lambda$, the strength of the RSOI, where the spin of excitations is locked to the direction of their momentum.  The electron excitations are denoted by solid circles and their horizontal arrows represent propagation directions whereas the vertical arrows is their spin directions. The vertical axis is the energy of excitations ($\varepsilon$) while the horizontal one is their momentum (${\bf k}$).}
   \label{Dispersion_relation}
 \end{figure}

When an $\uparrow$-spin electron from the left F electrode incident on the interface at $x=0$, not only there is a probability of $\uparrow$-spin electron reflection ($r_{N\uparrow}$), but also the probability of spin-flip reflection of electron ($r_{N\downarrow}$) is non-zero. This fact is a main consequence of the presence of RSOI region as a spin mixing barrier in the proposed junction.  The low energy band structures of the F1-RSO-F2 junction are shown in Fig.\ref{Dispersion_relation}.  In the F regions, the magnetization splits up the band structure into two sub-bands for $\uparrow$-spin and $\downarrow$-spin excitations. In the RSOI region, the spin and sudo-spin are coupled so that the spin-momentum locked bands are splited and we mark them by $\{\pm,\pm\}$ that refer to those $\pm$ appear in the eigenvalues of Eq. (\ref{EigenHSO}).

So, the wave functions in the three regions can be written as:

\begin{equation}
\begin{array}{cc}
\Psi_{\textnormal{F1}}=\psi_{e\uparrow}^{+}+r_{N\uparrow}\psi_{e\uparrow}^-+r_{N\downarrow}\psi_{e\downarrow}^-,
\\
\Psi_{\textnormal{RSO}}=a_1\psi_{\eta=+1}^{+}+a_2\psi_{\eta=+1}^{-}+a_3\psi_{\eta=-1}^{+}+a_4\psi_{\eta=-1}^{-},
\\
\Psi_{\textnormal{F2}}=t_{e\uparrow}\psi_{e\uparrow}^++t_{e\downarrow}\psi_{e\downarrow}^+,
\end{array}
\end{equation}

in which $t_{e\uparrow}$ and $t_{e\downarrow}$ refer to the probability of transmission electrons with $\uparrow$- and $\downarrow$-spin directions in the F2 region, respectively.
It should be noted that each $\psi_{e\uparrow(\downarrow)}$ has different values in F1 and F2 regions because of difference in the value of applied magnetic exchange field.
The wave functions associated with the dispersion relation in the F regions are:

  \begin{equation}
 \begin{array}{l}
\psi^{\pm}_{e\uparrow}(x)=\left(\mathbf{0^2}, 1 , \pm e^{\pm i
    \alpha_{e\uparrow}}, \mathbf{0^4}\right)^{\bf T} e^{\pm i
  k_{x\uparrow}^{\text{F}} x},\\
  
\psi^{\pm}_{e\downarrow}(x)=\left( 1 , \pm e^{\pm i \alpha_{e\downarrow}}, \mathbf{0^2},\mathbf{0^4}\right)^{\bf T} e^{\pm i k_{x\downarrow}^{\text{F}} x},
\\

  \end{array}
     \end{equation} 
where $\mathbf{0^n}$ represents a $1 \times n$ matrix with only zero entries and ${\bf T}$ is a transpose operator. We assume that the junction width $W$ is enough large so that the $y$ component of
the wave vector $k_y$ is a conserved quantity upon the scattering
processes and therefore, we factored out the corresponding
multiplication {\it i.e.} $\exp(i k_y y)$.

The momentum of the incident $\uparrow$-spin electron makes an angle $\alpha_{e\uparrow}$ with the $x$-axis. The angles of reflections inside the F1 barrier then given by,

\begin{equation}
   \alpha_{e\downarrow}=\arctan\left(\frac{q_n}{ k_{e\downarrow}^{x}} \right).
   \end{equation}
We denote $k_y^i\equiv q_n$ that can vary in interval $-\infty \leq
q_n \leq +\infty$. The $x$ component of the wavevector however becomes imaginary for values of $q_n$ larger than a critical value
$q^c$. The wavefunctions for $q_n>q^c$ are decaying functions and
therefore, depending on the junction geometry, are not able to contribute to the transport process. 

Translational invariance in the transverse (y) direction implies conservation of transverse momentum,

\begin{equation}
k_y=q_n \rightarrow k_y\sin\alpha_e=q_n\sin\alpha_e.
\end{equation}

The $x$-component of wavevectors are not conserved during the
scattering processes. So they express as:

   \begin{equation}
    \begin{cases}
   k_{x\uparrow}^{\text{F1}}=\left(\varepsilon+\mu+h_1\right)\cos\alpha_{e\uparrow},\\
    k_{x\downarrow}^{\text{F1}}=\left(\varepsilon+\mu-h_1\right)\cos\alpha_{e\downarrow},\\
    k_{x\uparrow}^{\text{F2}}=\left(\varepsilon+\mu+h_2\right)\cos\alpha_{e\uparrow},\\
    k_{x\downarrow}^{\text{F2}}=\left(\varepsilon+\mu-h_2\right)\cos\alpha_{e\downarrow},\\
    \end{cases}.
    \label{Eq.10}
    \end{equation}

in which $\hbar v_F=1$.
By matching the wave functions at the boundaries, \textit{i.e.},
$\Psi_{\textnormal{F1}}=\Psi_{\textnormal{RSO}}$ at $x=0$, and $\Psi_{\textnormal{RSO}}=\Psi_{\textnormal{F2}}$ at $x = L_{\textnormal{RSO}}$, we first calculate the transmission probabilities of charge carriers through graphene junction. Then, we perform the calculation of the conductance in the
tunneling junctions for the $\uparrow$-spin and $\downarrow$-spin electrons (See Ref.\cite{Beenakker2006}).

\begin{figure}
 \centering
   \begin{tabular}{c}
     \includegraphics[width=8cm, height=6cm]{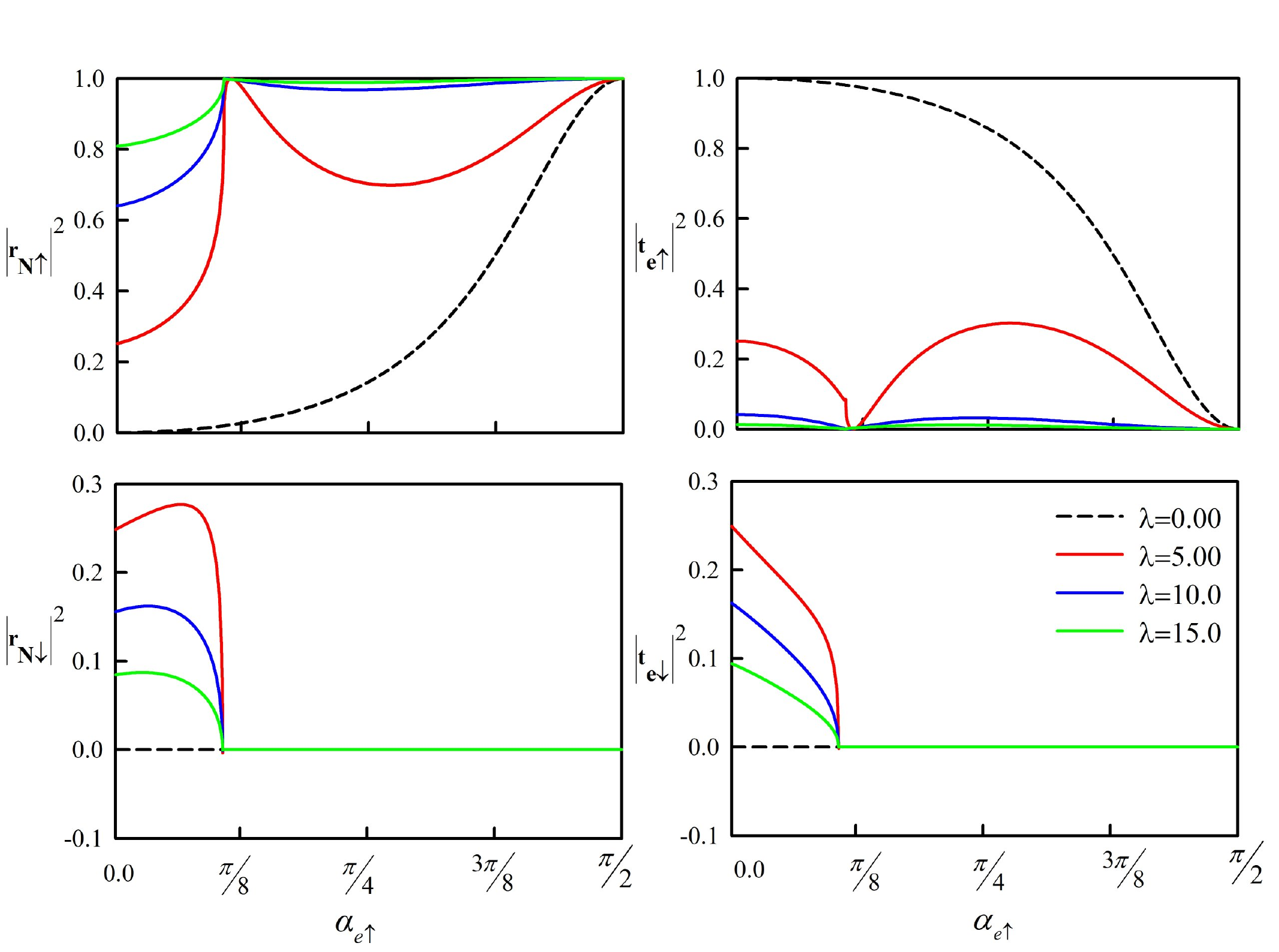}\\ 
   \end{tabular}
   \begin{tabular}{c}
        \includegraphics[width=8cm, height=6cm]{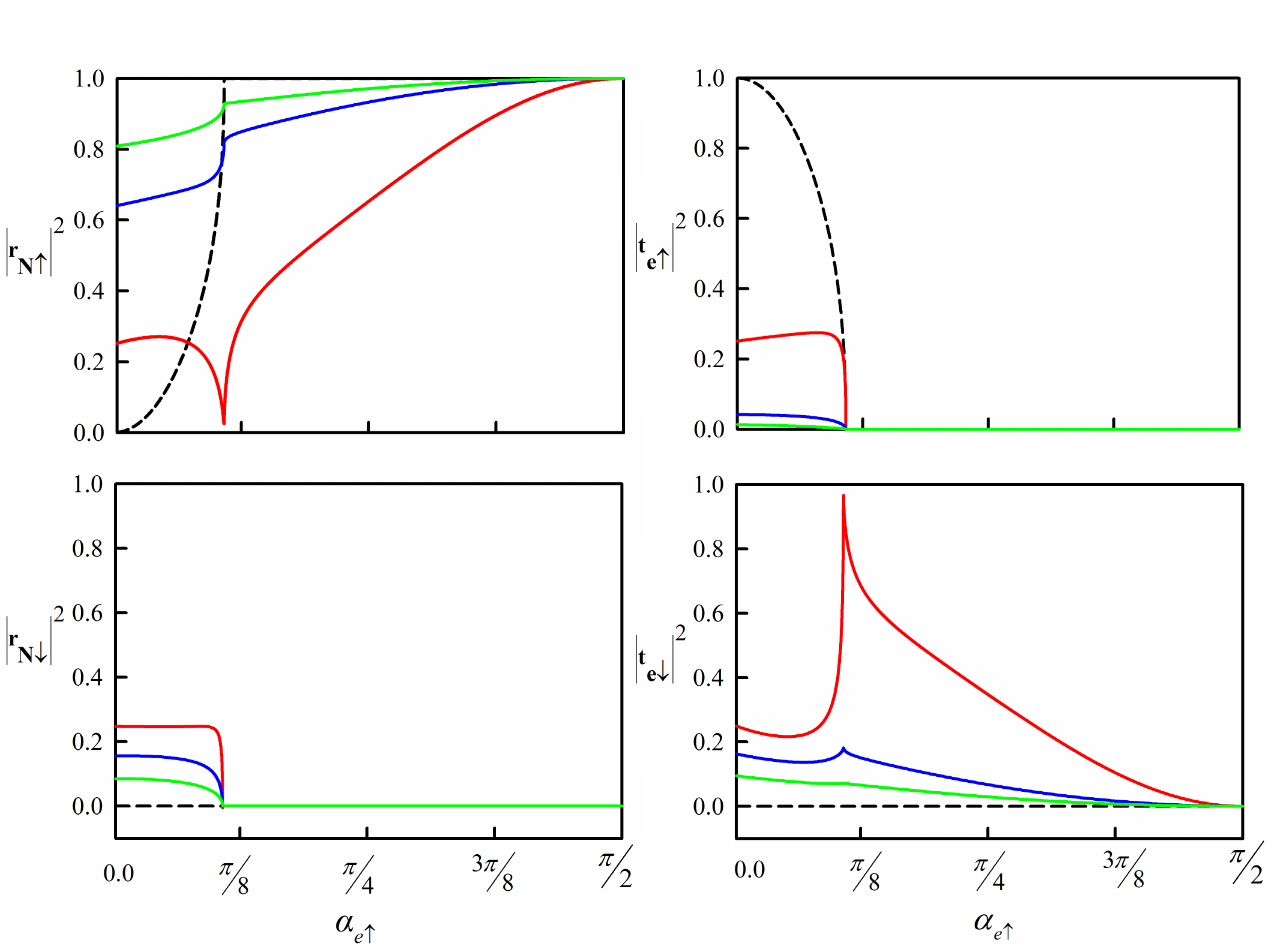}\\ 
      \end{tabular}
      \caption{Transmission and reflection probabilities through F1-RSO-F2 junction as a function of the incident angle for parallel configuration (Top panels) and anti-parallel configuration (bottom panels) in un-doped regime ($\mu= 0$). The parameters used in the calculation are $h_1=h_2=1$ and $\varepsilon=1$. The Rashba term ($\lambda$) varies from zero to 15.}  
   \label{Transmission_undoped}
 \end{figure}

\begin{figure}
 \centering
   \begin{tabular}{c}
     \includegraphics[width=8cm, height=6cm]{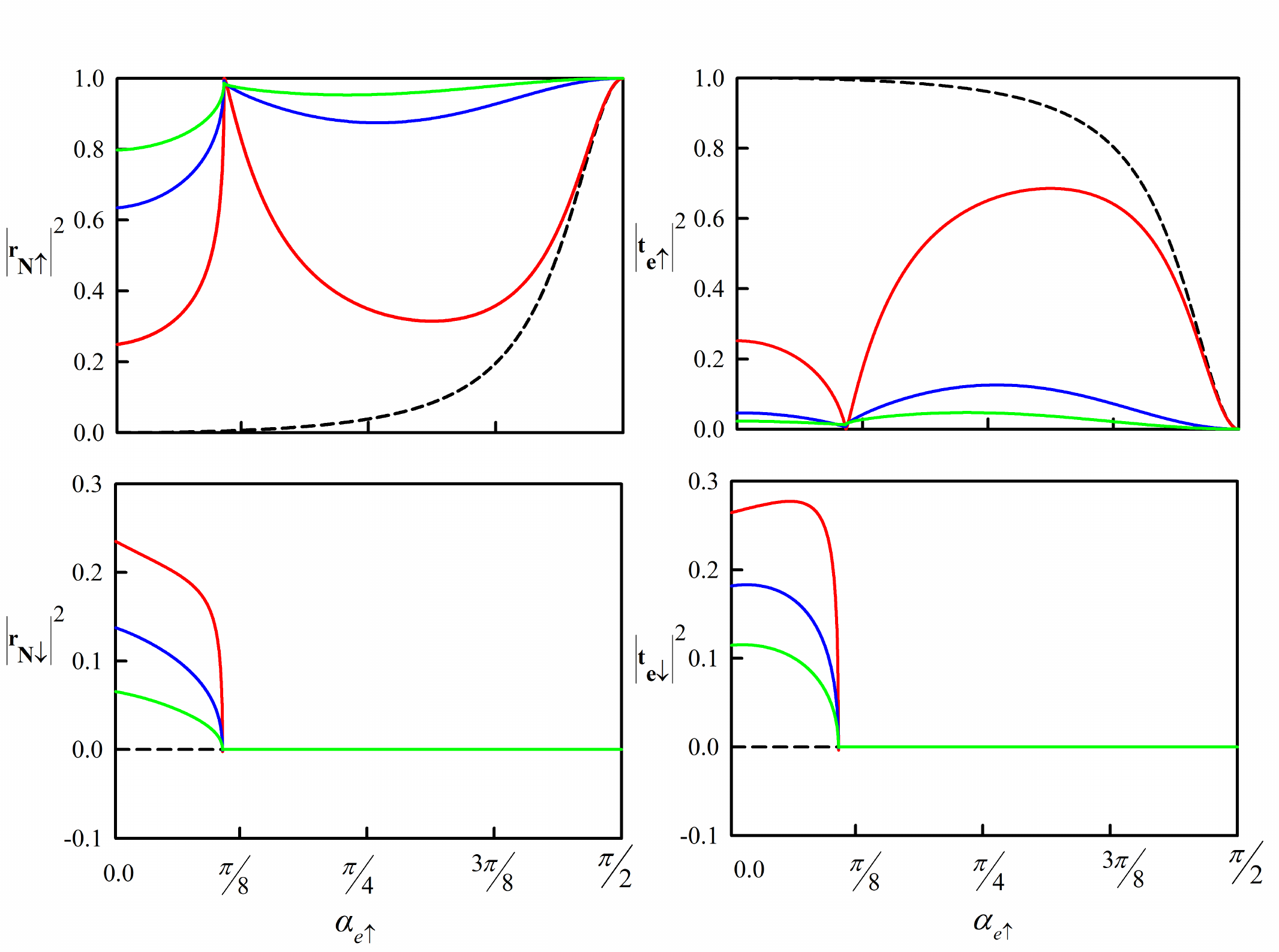}\\ 
   \end{tabular}
   \begin{tabular}{c}
        \includegraphics[width=8cm, height=6cm]{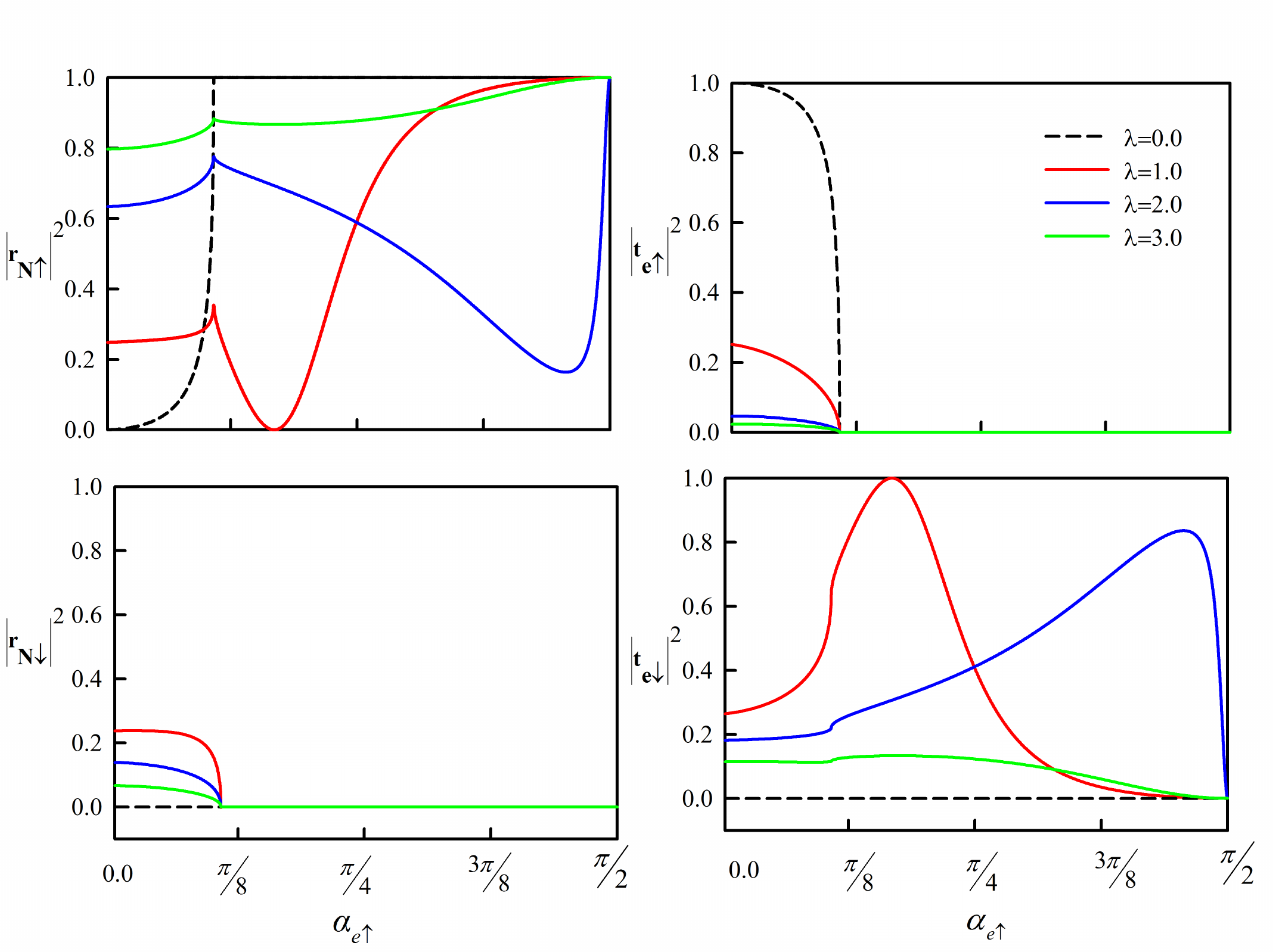}\\ 
      \end{tabular}
      \caption{Same as Fig. \ref{Transmission_undoped}, but for $\mu\neq 0$.}   
   \label{Transmission_doped}
 \end{figure}

 The transmission probabilities of $\uparrow$-spin electrons as a function of the incident angle ($\alpha_{e\uparrow}$) are presented in Figs. \ref{Transmission_undoped} and \ref{Transmission_doped}. Since the Fermi energy of the graphene material can be tuned by the local chemical doping \cite{Nov2004}, we can consider two different regimes named un-doped ($\mu\approx 0$) and doped ($\mu\neq 0$) graphene. The chemical potential is vanishingly small in an un-doped graphene sheet. 

It should be noted that, in the graphene-based junctions, unlike spin–orbit splitting in the conventional materials, which is usually small compared to the Fermi energy (it is about 10-11 meV), the strength of RSOI may be comparable to or even be bigger than the Fermi energy of the electrons. Because, the RSOI couples the pseudo- and the real-spins in graphene. So we can set $\lambda$ up to 120 meV, which corresponds to the typical Fermi energy of the electrons in graphene material. Thus the parameters used in the following diagrams are realistic and the results are exact.

In the absence of RSOI, the transmission spectrum for P configuration shows angular anisotropy and Klein tunneling, obviously. Due to the presence of RSOI, Klein tunneling is blocked and a new type of normal reflection can also take place which we called it, spin-flip normal reflection (See Figs. \ref{Transmission_undoped} and \ref{Transmission_doped}). The scattering probabilities can be tuned via doping ($\mu$), applied magnetic exchange field in each ferromagnetic electrodes ($h_1$ and $h_2$) and even the strength of Rashba term ($\lambda$). 
For AP configuration, perfect transmission (Klein tunneling) can not be occurred because the state of carriers in the
two ferromagnetic electrodes are not the same(except for normal incidence with $\alpha_\uparrow=0$). However, there is a probability for spin-flipped Klein tunneling at oblique incident angle in the presence of RSOI.

The dispersion relation diagram of the F1-RSO-F2 junction (Fig. \ref{Dispersion_relation}) shows that the RSOI opens a gap in the energy spectrum, so it has a great impact on transport
properties of quasi-particles in the junction. These also are discussed in some previous studies by a different way \cite{Bai2010,Liu2012}. 
Another significant feature of such a junction is the ability to split the spin current in both sides of junction. Let us consider that a $\uparrow$-spin electron hit the interface at $x=0$. It can reflect back as a $\uparrow$-spin electron or even $\downarrow$-spin one. 
We can easily tune the input parameters in the junction to reach different regimes in which the probability of spin-flip normal reflection (or transmission) is dominant while the probability of normal reflection (or transmission) is negligible. In such a regime, there is an input current with $\uparrow$-spin carriers in F1 electrode which produced a $\downarrow$-spin output current in F2 electrode. This effect is obviously demonstrated in Figs. \ref{Transmission_undoped} and \ref{Transmission_doped}.
So, we can suggest this experimentally feasible junction as an spin-valve for spintronic aims. In experimental setups, spin (angle)-resolved photo-emission spectroscopy was employed to observe the spin-dependent band splitting in the junctions \cite{Varykhalov2008,Dil2009,Meier2009}.

So far, we have only discussed on the angle-resolved transmission probabilities. Since the conductance is more accessible in experiments than the transmission probabilities, we now turn to a discussion on the charge and spin conductances.
Based on the Landauer-B\"uttiker formalism, let us define the
normalized charge and spin conductances for the different
magnetization configurations at zero temperature as,

\begin{equation}
\begin{array}{cc}
G_{ch}=\dfrac{G_{\uparrow}+G_{\downarrow}}{2},\\
\\
G_{sp}=\dfrac{G_{\uparrow}^s+G_{\downarrow}^s}{2},\\
\end{array}
\end{equation}

\begin{equation}
G_{\uparrow(\downarrow)}=g_0\int_{-\pi/2}^{\pi/2} \cos\alpha d\alpha   
\Big(1-\sum_{\uparrow(\downarrow)} 
(|r_{N\uparrow}|^2+|r_{N\downarrow}|^2 )\Big),
\end{equation}

\begin{equation}
G_{\uparrow(\downarrow)}^{s}=g_0\int_{-\pi/2}^{\pi/2} \cos\alpha d\alpha   
\Big(1-\sum_{\uparrow(\downarrow)} 
(|r_{N\uparrow}|^2-|r_{N\downarrow}|^2 )\Big),
\end{equation}

where  $g_0=N_\sigma{2e^2}/{h}$ represents the spin-dependent ballistic conductance of the junction as a function of density of state $N_\sigma=|\varepsilon+\mu+\sigma h|$ where $\sigma=\pm 1$. 

For both P and AP configurations, the total charge (spin) conductance across the junction is the sum of the two spin-dependent charge (spin) conductances. They are $G_{ch}^\textnormal{P} = G_{ch}^{\uparrow\uparrow} + G_{ch}^{\downarrow\downarrow}$ ($G_{sp}^\textnormal{P} = G_{sp}^{\uparrow\uparrow} + G_{sp}^{\downarrow\downarrow}$) for P and $G_{ch}^\textnormal{AP} = G_{ch}^{\uparrow\downarrow} + G_{ch}^{\downarrow\uparrow}$ ($G_{sp}^\textnormal{AP} = G_{sp}^{\uparrow\downarrow} + G_{sp}^{\downarrow\uparrow}$) for AP configurations. 

Then we obtain the charge and spin magnetoresistance (MR and SMR) as follow,

\begin{equation}
\begin{array}{cc}
\textnormal{MR}=(G_{ch}^\textnormal{P}-G_{ch}^\textnormal{AP})/G_{ch}^\textnormal{P},\\
\textnormal{SMR}=(G_{sp}^\textnormal{P}-G_{sp}^\textnormal{AP})/G_{sp}^\textnormal{P}.\\
\end{array}
\end{equation}

Since $G_{\uparrow\uparrow}=G_{\downarrow\downarrow}$ and $G_{\uparrow\downarrow}=G_{\downarrow\uparrow}$, MR and SMR can be simplified to $\textnormal{MR}=(G_{ch}^{\uparrow\uparrow}-G_{ch}^{\uparrow\downarrow})/G_{ch}^{\uparrow\downarrow}$ and $\textnormal{SMR}=(G_{sp}^{\uparrow\uparrow}-G_{sp}^{\uparrow\downarrow})/G_{sp}^{\uparrow\downarrow}$ , respectively.
We also introduce the spin polarization of the current
through the junction,  $\eta_P=(G_{\uparrow\uparrow}-G_{\downarrow\downarrow})/(G_{\uparrow\uparrow} + G_{\downarrow\downarrow})$ and $\eta_{AP}=(G_{\uparrow\downarrow}-G_{\downarrow\uparrow})/(G_{\uparrow\downarrow}+G_{\downarrow\uparrow})$ for the
P and AP configurations, respectively. 

\begin{figure}
 \centering
   \begin{tabular}{cc}
      \includegraphics[width=5cm, height=10cm]{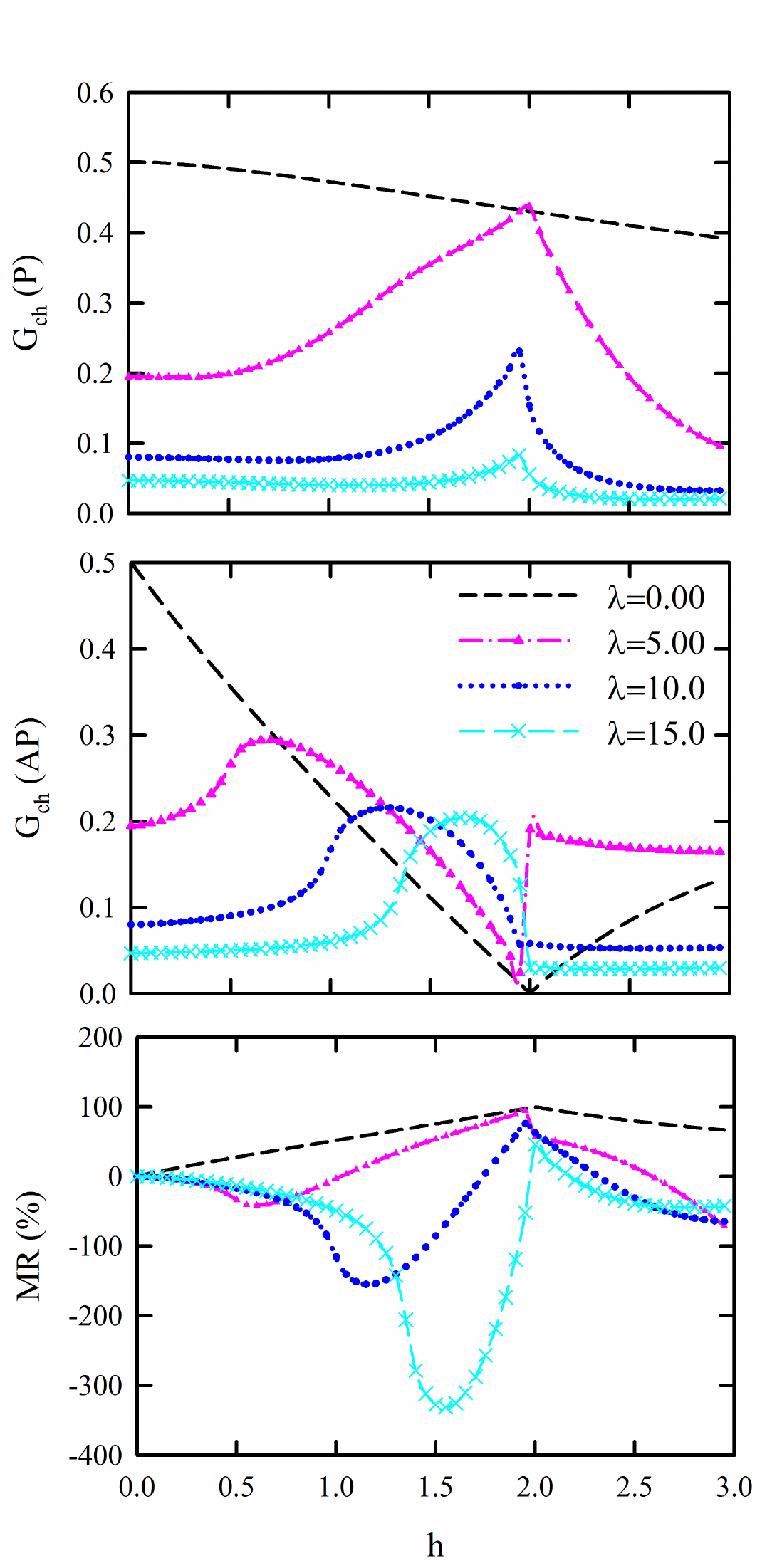}
        \includegraphics[width=5cm, height=10cm]{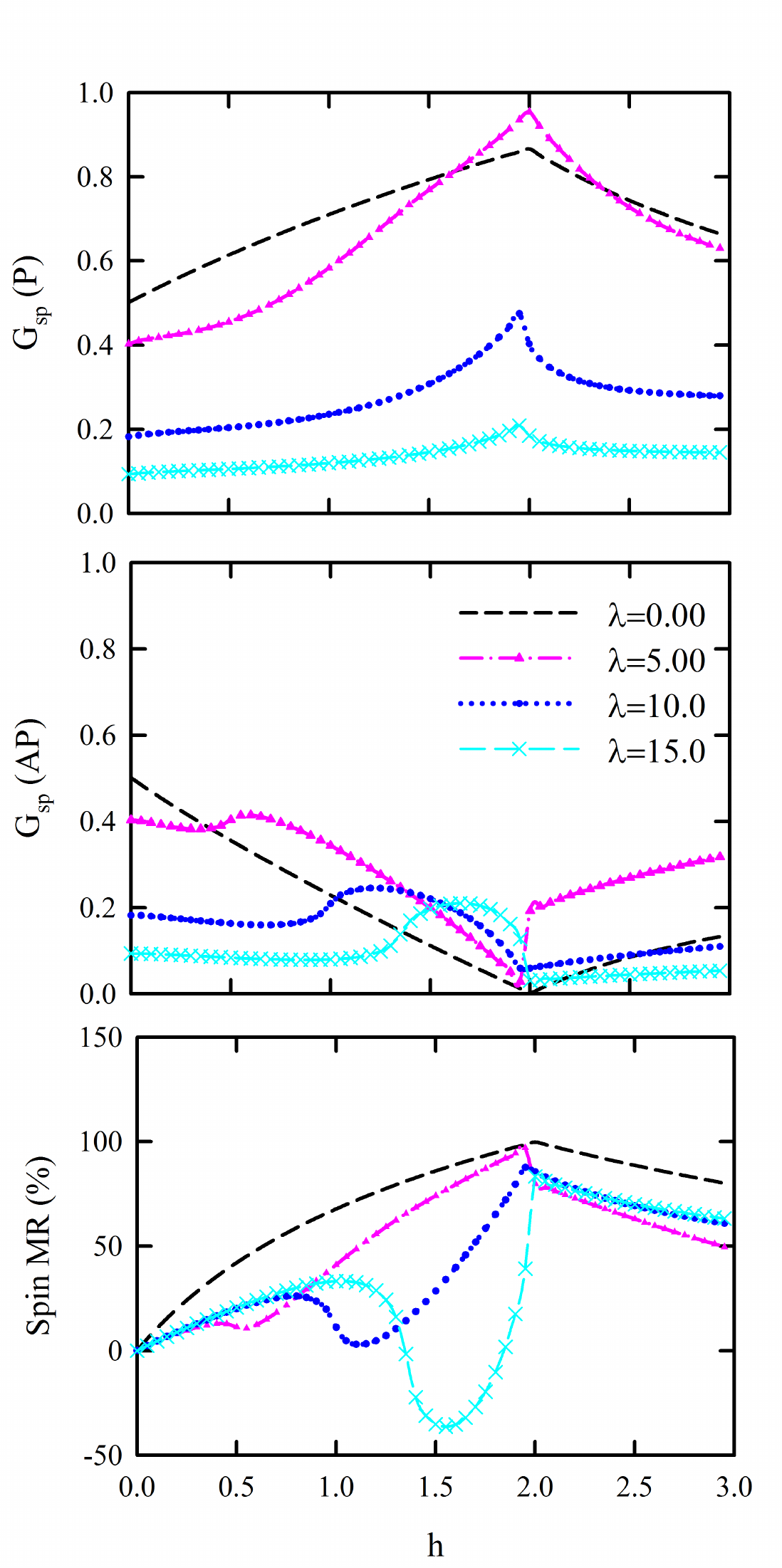} 
      \end{tabular}
      \caption{(Color online) The charge and spin conductances along with MR and SMR percentages as a function of normalized exchange field ($h$) are shown for both P and AP configurations. The magnitude of RSOI changes from zero to 15 and $\mu=\varepsilon=1$. }   
   \label{MR vs h}
 \end{figure}

 In Fig. \ref{MR vs h}, the charge and spin conductances along with MR and SMR for both P and AP configurations are shown as a function of $h$. The $x$-axis in this figure is the strength of magnetic exchange field. For simplicity, the strength of applied magnetic exchange field in two ferromagnetic electrode are supposed to be equal ($h_1=h_2$) and the value of RSOI parameter varies from zero to 15. 
 In the presence of RSOI, the sign of MR and SMR can switch from a positive value to a negative one. Also, the maximum value of MR can reach to 350 \% which is almost ten times larger than that in the previous calculation \cite{Bai2010,Liu2005}. 
 When the spin-orbit coupling is off ($\lambda=0$), the middle region acts as a flawless normal graphene sheet.  It can be seen from the Fig. \ref{MR vs h} that, in general, the MR ratio is small in the absence of RSOI.  But, when RSOI blocks one of the spin channels for carries, the MR ratio reaches to its maximum value up to 300\%. So, in order to get a high-value of the MR ratio, it is needed to shift between the center of the spin bands in the RSOI region.

\begin{figure}
 \centering
   \begin{tabular}{cc}
      \includegraphics[width=5cm, height=10cm]{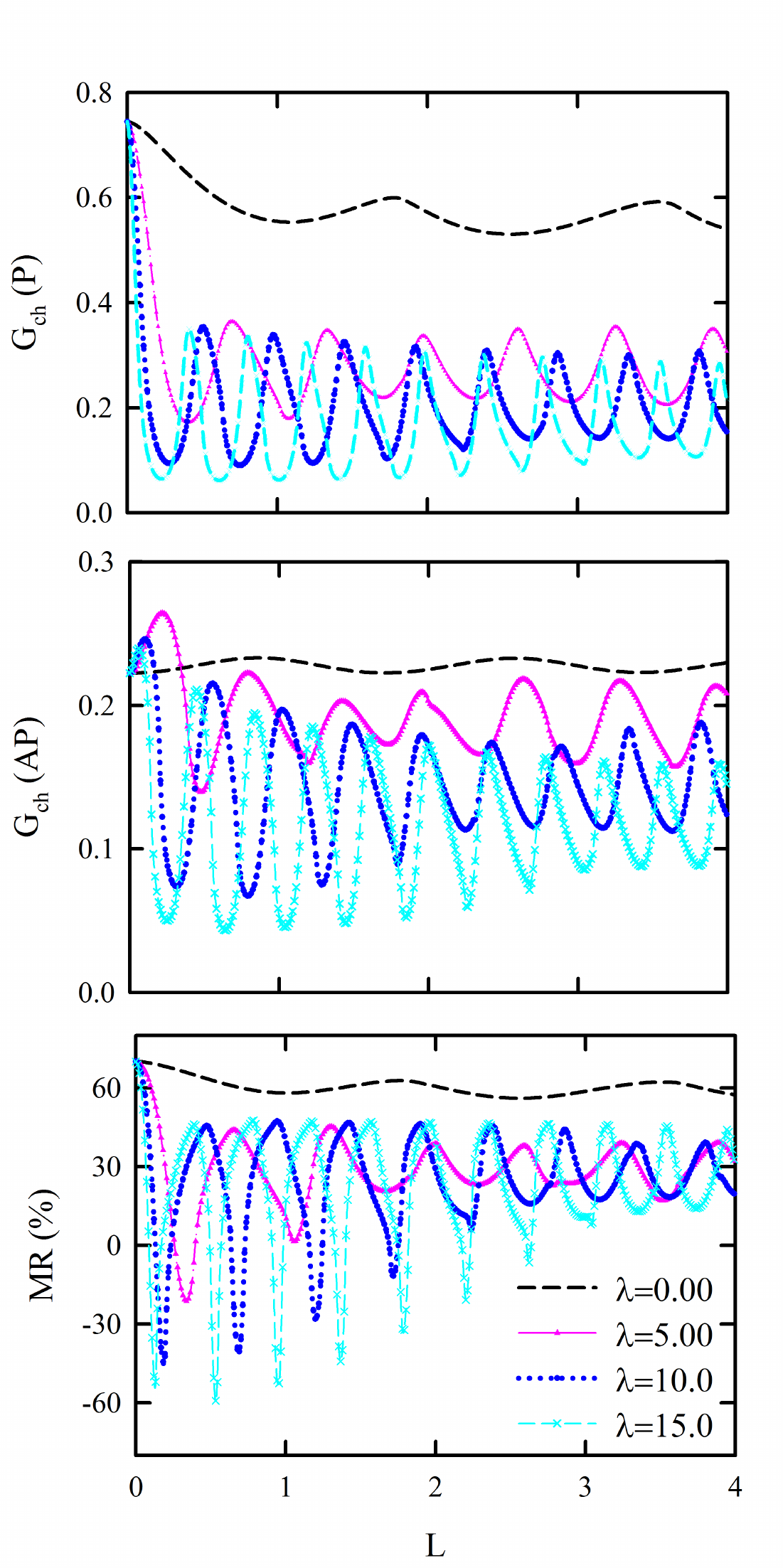}
        \includegraphics[width=5cm, height=10cm]{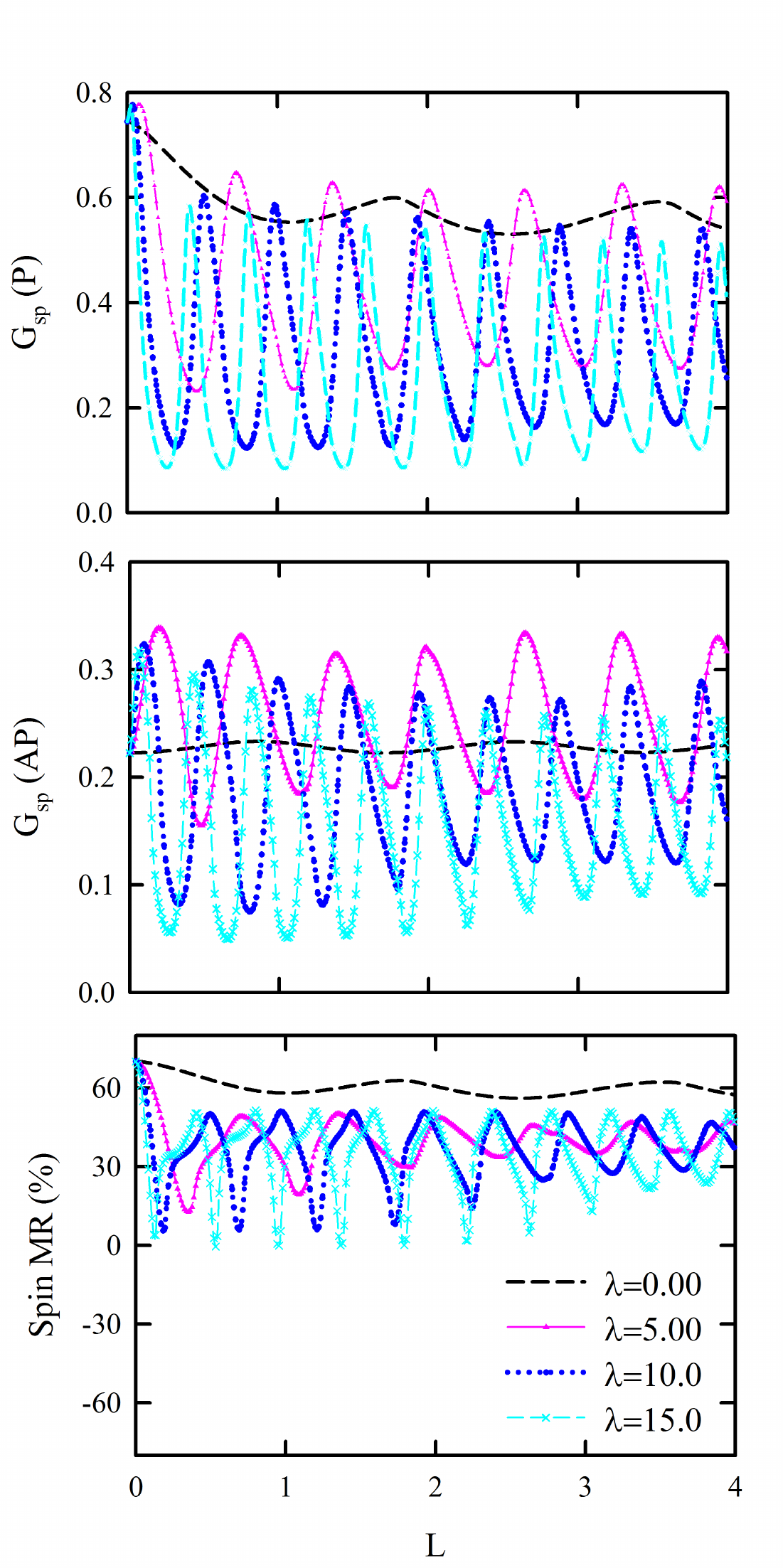} 
      \end{tabular}
      \caption{(Color online) The charge and spin conductances along with MR and SMR percentages as a function of length ($L$) for both P and AP configurations. The magnitude of RSOI changes from zero to 15 and $\mu=\varepsilon=1$. }
   
   \label{MR vs L}
 \end{figure}

In the next step, we show the dependence of MR and SMR to the length of the junction for both P and AP configurations. Fig. \ref{MR vs L} show the charge and spin conductances accompany with the MR and SMR ratios as a function of length (L). It can be seen from Fig. \ref{MR vs L} that, due to the spin-dependent Klein tunneling, both MRs and conductances exhibit the oscillation features.
The results show that the P and AP conductances have the same oscillation periods.
At high strength of RSOI parameter, the frequency of oscillation increases. This effect originate from the influence of RSOI region as a spin-mixer barrier. The RSOI region rotates the spin direction of the carriers. So, the final spin direction of the carriers in F2 region depend on the strength of RSOI as well as the length of the junction.

Because of the particle-wave property of electrons, there is a current fluctuations out of equilibrium which is called shot-noise \cite{Jong1996}. The density-dependent shot-noise in graphene and the effect of disorders on it have been studied theoretically  before \cite{Tworzydlo2006,DiCarlo2008,Danneau2008}. In our wave-function scattering procedure, the shot noise of the system has the following form,

\begin{equation}
S=\int_{-\pi/2}^{\pi/2} T(1-T)\cos\alpha d\alpha, 
\end{equation}

where T is the total transmission probabilities in F2 electrode. 
In this section we calculate the shot-noise in graphene-based junction with and without RSOI. Next, we calculate the ratio of the actual shot-noise and the Poissonian noise that is called Fano factor ($F$) for both P and AP configurations,

\begin{equation}
\begin{array}{cc}
F^P=\dfrac{\int_{-\pi/2}^{\pi/2} T^P(1-T^P)\cos\alpha d\alpha}{G^P},\\
\\
F^{AP}=\dfrac{\int_{-\pi/2}^{\pi/2} T^{AP}(1-T^{AP})\cos\alpha d\alpha}{G^{AP}},\\
\end{array}
\end{equation}

where $T^P=T_{\uparrow\uparrow}=T_{\downarrow\downarrow}$ and $T^{AP}=T{\uparrow\downarrow}=T_{\downarrow\uparrow}$.

\begin{figure}
 \centering
      \includegraphics[width=10cm, height=8cm]{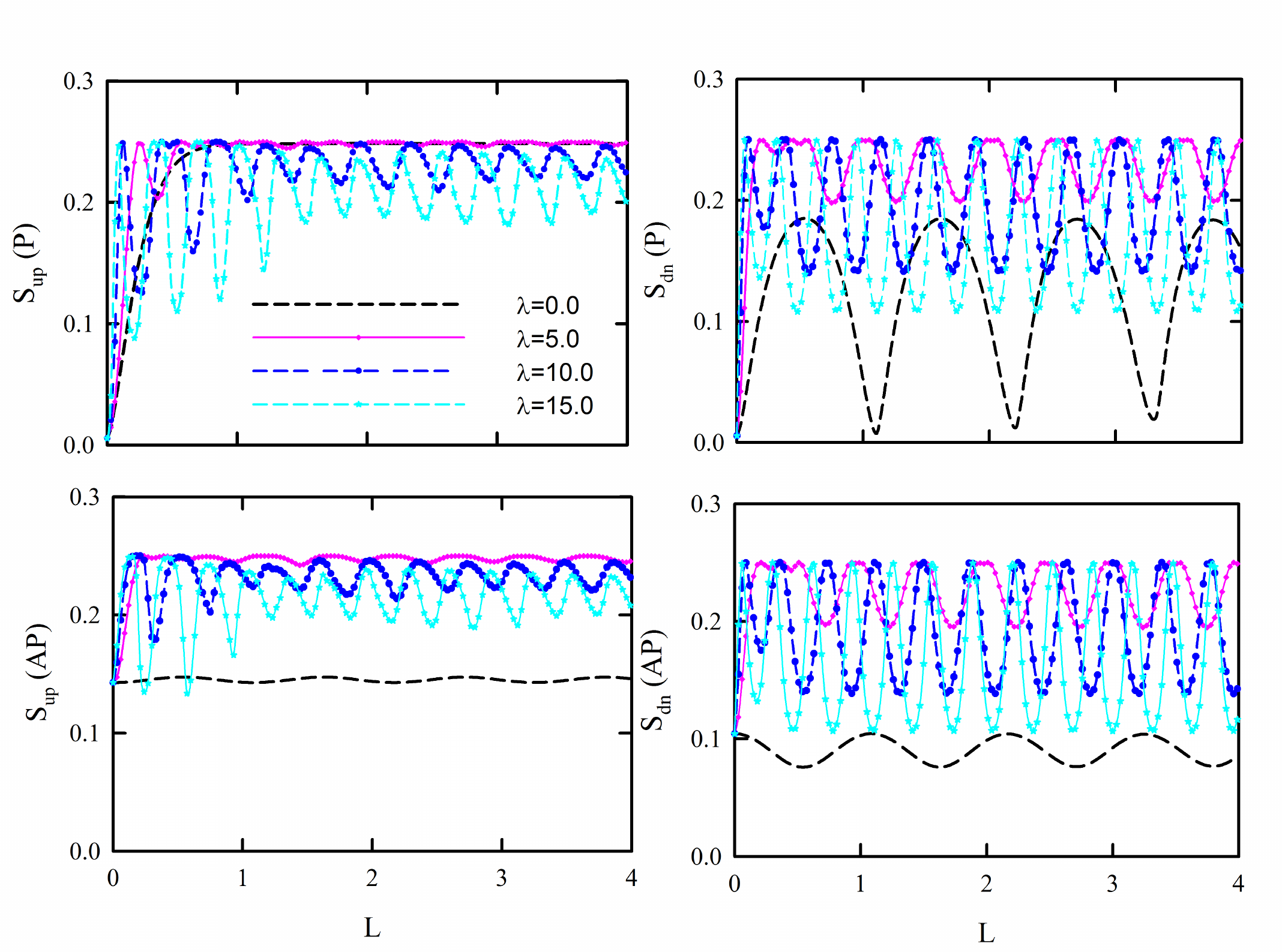}
      \caption{(Color online) Shot noise as a function of length ($L$) for both P and AP configurations. The parameters used in the calculation are $h_1=h_2=2$, $\mu=2$ and $\varepsilon=1$. The magnitude of RSOI changes from zero to 15. }
   \label{S_vs_L}
 \end{figure}
 
In Fig. \ref{S_vs_L}, we present the typical dependence of the shot-noise on the length of the junction. 
As it can be seen from Fig. \ref{S_vs_L}, the parallel and anti-parallel shot-noises show oscillatory behavior as expected. But the amplitude and frequency of these oscillations increased with increasing the strength of Rashba term. 
The phenomena of frequency change originates from
the interference effects between the quasi-particle's wave functions in the RSO region. That is to say, when the strength of RSOI is large, the amplitude and frequency of oscillations are increased because the structure turn into a more complex structure. 
On the other hand, as the strength of RSOI increases, the value of AP conductance can become larger than that of P configuration due to the spin-mixing effect. 
So, the P and AP shot-noises do not behave as a monotonic function with increasing the RSOI strength, as seen in Fig.\ref{S_vs_L}.

\begin{figure}
 \centering
      \includegraphics[width=12cm, height=4cm]{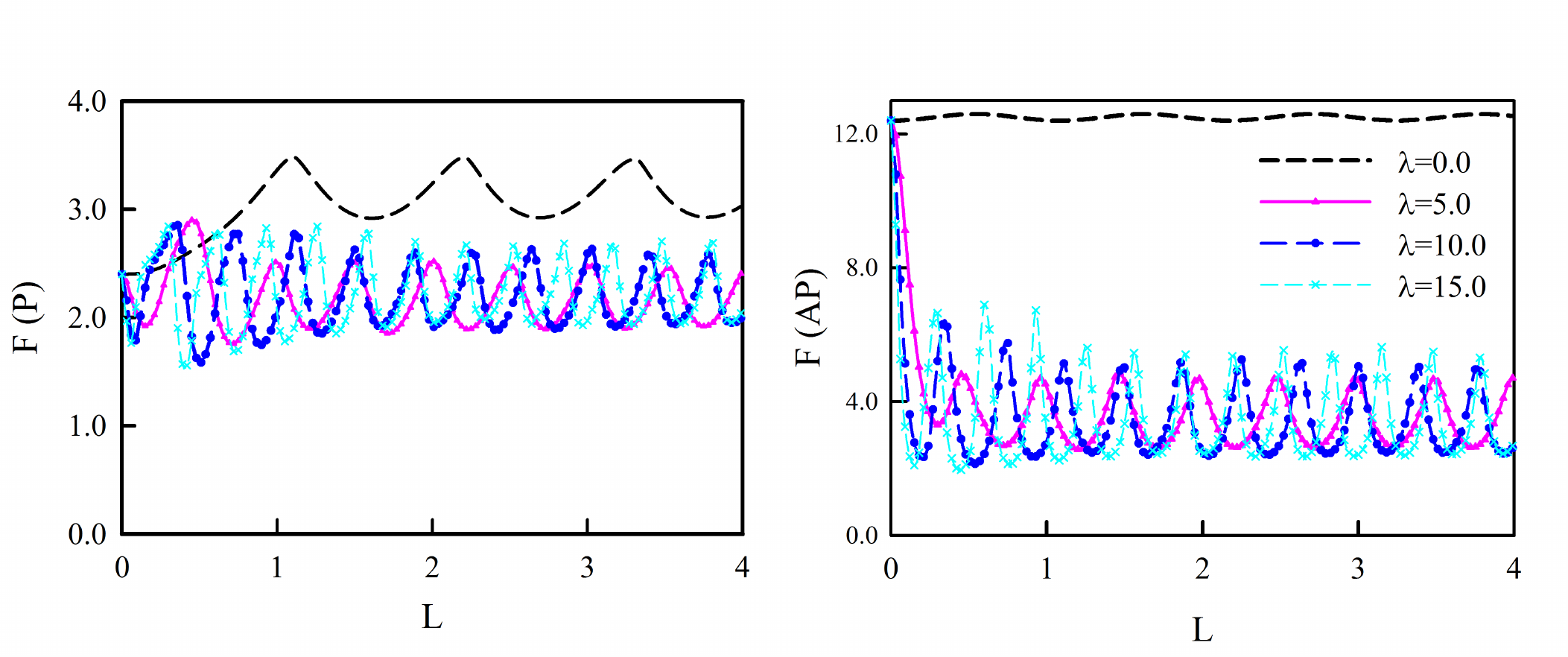}

      \caption{(Color online) Fano factor as a function of length ($L$) for both P and AP configurations. The parameters used in the calculation are $h_1=h_2=2$, $\mu=2$ and $\varepsilon=1$. The magnitude of RSOI changes from zero to 15. }
   
   \label{Fano_vs_L}
 \end{figure}

Next we study the Fano factor diagram of such a structure. 
The total Fano factor in this system is shown in Fig. \ref{Fano_vs_L} for both P and AP configurations.
The incident energy and total length are the same as those in Fig. \ref{S_vs_L}.

In contrast to the behavior seen in the absence of RSOI, we do not observe a clear scaling behavior for Fano factor in the presence of RSOI, specially in AP case. By the result of Fig. \ref{Fano_vs_L}, the oscillation feature of Fano factor is
changed due to the increasing of the RSOI strength.
As known, the universal maximum value of 1/3 for Fano factor in graphene increased with decreasing in the density of charge carriers \cite{Tworzydlo2006}. Since the RSOI change the density of carriers, we can control the position and amplitude of the peaks by applying different values of $\lambda$.

\section{Conclusion}
In summary, we have shown how the angle resolved transmission probability in the proposed structures may be tuned by the RSOI strength. 
We also show that the spin-dependent conductance strongly depends on the strength of the RSOI. This conductance is a sensitive oscillatory function of the thickness of the RSO region. Because of the spin-flip effect, the junction shows a spin-valve effect with large and negative magnetoresistance (MR) and spin-magnetoresistance (SMR) in the presence of RSOI. We also study the variation of shot-noise and Fano factor due to the RSOI. Results show that the Fano factor can be tuned largely by the magnetic exchange field and RSOI.
As the graphene material is more flexible and simple in design than other nanoscale 2D materials, this proposed structures may have a more expectance compared to the conventional semiconductor junctions.
\\
\\
{\bf Acknowledgment}
\\

The authors acknowledge the Iran national science foundation (INSF) for financial support.
\\


{\bf References}
\begin{thebibliography}{000}


\bibitem{DasSarma2001} S. Das Sarma, J. Fabian, X. Hu, I. Zuti\'{c}, {\it Spin electronics and spin computation}, \href{URL}{Sol. St. Comm., {\bf 119}, 207 (2001)}.

\bibitem{Zutic2004}  I. \'{Z}uti\'{c}, J. Fabian, S. Das Sarma, {\it Spintronics: Fundamentals and applications}, \href{URL}{Rev. Mod. Phys, {\bf 76}, 323 (2004)}.


\bibitem{Kovalev2006}A.A. Kovalev, G.E.W. Bauer, A. Brataas, {\it  Perpendicular spin valves with ultrathin ferromagnetic layers: Magnetoelectronic circuit investigation of finite-size effects}, \href{URL}{Phys. Rev. B {\bf 73} 054407 (2006)}.

\bibitem{Brataas2006}A. Brataas, Y. Tserkovnyak, G.E.W. Bauer, {\it Current-induced macrospin versus spin-wave excitations in spin valves},\href{URL}{Phys. Rev. B {\bf 73}, 014408 (2006)}.


\bibitem{Hill2006} E.W. Hill, A.K. Geim, K. Novoselov, F. Schedin, P. Blake, {\it Graphene Spin Valve Devices}, \href{URL}{ Mag., IEEE Trans., {\bf 42}, 2694 (2006)}.

\bibitem{Zhai2008} F. Zhai, and K. Chang, {\it Theory of huge tunneling magnetoresistance in graphene}, \href{https://journals.aps.org/prb/abstract/10.1103/PhysRevB.77.113409}{Phys. Rev. B {\bf 77}, 113409 (2008)}.
    
 
 \bibitem{Brey2007} L. Brey, and H. A. Fertig, {\it Magnetoresistance of graphene-based spin valves}, \href{https://journals.aps.org/prb/abstract/10.1103/PhysRevB.76.205435}{Phys. Rev. B {\bf 76}, 205435 (2007)}.
 
 \bibitem{Rojas2009} F. M-Rojas, J. F-Rossier, and J. J. Palacios, {\it Giant Magnetoresistance in Ultrasmall Graphene Based Devices}, \href{https://journals.aps.org/prl/abstract/10.1103/PhysRevLett.102.136810}{Phys. Rev. Lett. {\bf 102}, 136810 (2009)}.
    
\bibitem{Bai2010} C. Bai, J. Wang, S. Jia, and Y. Yang, {\it Spin-orbit interaction effects on magnetoresistance in graphene-based ferromagnetic double junction}, \href{http://aip.scitation.org/doi/10.1063/1.3432438}{App. Phys. Lett. {\bf 96}, 223102 (2010)}.   

\bibitem{Varykhalov2008} A. Varykhalov, J. Snchez-Barriga, A. M. Shikin, C. Biswas, E. Vescovo, A. Rybkin, D. Marchenko, and O. Rader, {\it Electronic and Magnetic Properties of Quasifreestanding Graphene on Ni} \href{https://journals.aps.org/prl/abstract/10.1103/PhysRevLett.101.157601}{Phys. Rev. Lett. {\bf 101}, 157601
(2008)}.

\bibitem{Dil2009} J. H. Dil, {\it Spin and angle resolved photoemission on non-magnetic low-dimensional systems} \href{http://iopscience.iop.org/article/10.1088/0953-8984/21/40/403001/meta;jsessionid=07CDBC22EE5656BC01D1A8B81E61E3BC.c4.iopscience.cld.iop.org}{J. Phys.: Condens. Matter {\bf 21}, 403001 (2009)}. 

\bibitem{Meier2009}F. Meier, J. H. Dil, and J. Osterwalder, {\it Measuring spin polarization vectors in angle-resolved photoemission spectroscopy} \href{http://iopscience.iop.org/article/10.1088/1367-2630/11/12/125008/meta}{New J. Phys. 11, 125008 (2009)}. 

\bibitem{Wu2011} Q-P. Wu, X-D. H, and Z-F. Liu, {\it A wide-angle spin filter based on graphene with Rashba coupling and exchange field}, \href{http://www.sciencedirect.com/science/article/pii/S138694771100422X?via}{Physica E {\bf 77}, 738 (2011)}.

\bibitem{Liu2012} Z-F. Liu, N-H. Liu, and Q-P. Wu, {\it Magnetoresistance and shot noise in graphene-based nanostructure with effective exchange field}, \href{http://aip.scitation.org/doi/10.1063/1.4770494}{J. App. Phys. {\bf 112}, 123719 (2012)}.                 


  \bibitem{Wang2015} Z. Wang, C. Tang, R. Sachs, Y. Barlas, J. Shi,
     {\it Proximity-induced ferromagnetism in graphene revealed by the anomalous Hall effect}, \href{http://journals.aps.org/prl/abstract/10.1103/PhysRevLett.114.016603}{Phys. Rev. Lett {\bf 114}, 016603 (2015)}.
         
     \bibitem{Avsar2014} A. Avsar, J. Y. Tan, T. Taychatanapat, J. Balakrishnan, G.K.W. Koon, Y. Yeo, J. Lahiri, A. Carvalho, A. S. Rodin, E.C.T. O’Farrell, G. Eda, A. H. Castro Neto and B. Ozyilmaz, {\it Spin orbit proximity effect in graphene},
     \href{http://www.nature.com/ncomms/2014/140926/ncomms5875/full/ncomms5875.html?WT.ec_id=NCOMMS-20141001}{Nat.Commun., 
     {\bf 5}, 4875 (2014)}. 
     

                     
\bibitem{Nov2004} K. S. Novoselov, A. K. Geim, S. V. Morozov, D. Jiang, Y. Zhang, S. V. Dubonos, I. V. Grigorieva, and A. A. Firsov, {\it Electric Field Effect in Atomically Thin Carbon Films}, \href{http://science.sciencemag.org/content/306/5696/666}{Science {\bf 306}, 666 (2004)}.

\bibitem{Nov2005} K. S. Novoselov, A. K. Geim, S. V. Morozov, D. Jiang, M. I. Katsnelson, I. V. Grigorieva, S. V. Dubonos, and A. A. Firsov, {\it Two-dimensional gas of massless Dirac fermions in graphene}, 
\href{http://www.nature.com/nature/journal/v438/n7065/abs/nature04233.html}{Nature {\bf 438}, 197 (2005)}.

\bibitem{Park2008} C. Park, L. Yang, Y. Son, M. L. Cohen, and S. G. Louie, {\it Anisotropic behaviors of massless Dirac fermions in graphene under periodic potentials}, \href{http://www.nature.com/nphys/journal/v4/n3/full/nphys890.html}{ Nat. Phys. {\bf 4}, 213 (2008)}.

\bibitem{Nov2009} A. H. Castro Neto, F. Guinea, N. M. R. Peres, K. S. Novoselov, and A. K. Geim, {\it The electronic properties of graphene}, \href{http://journals.aps.org/rmp/abstract/10.1103/RevModPhys.81.109}{Rev. Mod. Phys. {\bf 81}, 109 (2009)}.

\bibitem{Bonaccorso2010} F. Bonaccorso, Z. Sun, T. Hasan, and A. C. Ferrari, {\it Graphene photonics and optoelectronics}, 
\href{http://www.nature.com/nphoton/journal/v4/n9/full/nphoton.2010.186.html}{Nat. Pho. {\bf 4}, 611 (2010)}.

\bibitem{Avouris2010}P. Avouris, {\it Graphene: Electronic and photonic properties and devices}, \href{http://pubs.acs.org/doi/abs/10.1021/nl102824h}{Nano. Lett. 
{\bf 10} 4285 (2010)}.

\bibitem{Bao2012} Q. Bao, and K. P. Loh, {\it Graphene photonics, plasmonics, and broadband optoelectronic devices}, \href{http://pubs.acs.org/doi/abs/10.1021/nn300989g}{ACS Nano, {\bf 6}, 3677 (2012)}.

\bibitem{Zhang2005} Y. Zhang, Y-W. Tan, H. L. Stormer, and P. Kim, {\it Experimental observation of the quantum Hall effect and Berry's phase in graphene}, \href{https://www.nature.com/nature/journal/v438/n7065/full/nature04235.html}{Nature {\bf 438}, 201 (2005)}.

\bibitem{Son2006} Y-W. Son, M. L. Cohen, S. G. Louie, {\it Half-Metallic Graphene Nanoribbons}, \href{https://www.nature.com/nature/journal/v444/n7117/full/nature05180.html}{Nature {\bf 444} 347 (2006)}.

\bibitem{Nair2012} R. R. Nair, M. Sepioni, I-L. Tsai, O. Lehtinen, J. Keinonen, A. V. Krasheninnikov, T. Thomson, A. K. Geim, I. V. Grigorieva, {\it Spin-half paramagnetism in graphene induced by point defects}, \href{http://www.nature.com/nphys/journal/v8/n3/full/nphys2183.html}{Nat. Phys. {\bf 8}, 199 (2012)}.

\bibitem{Pesin2012}D. Pesin, and A. H. MacDonald, {\it Spintronics and 
pseudospintronics in graphene and topological insulators}, 
\href{http://www.nature.com/nmat/journal/v11/n5/abs/nmat3305.html}{Nat. Mat. 
{\bf 11}, 409 (2012)}.


\bibitem{Han2014} W. Han,R. K. Kawakami, M. Gmitra, and J. Fabian, {\it Graphene spintronics}, 
\href{http://www.nature.com/nnano/journal/v9/n10/abs/nnano.2014.214.html}{Nat. 
Nanotech. {\bf 9}, 807 (2014)}.

\bibitem{Zareyan2010} A. G. Moghaddam, and M. Zareyan, {\it Graphene-based electronic spin lenses}, 
\href{http://journals.aps.org/prl/abstract/10.1103/PhysRevLett.105.146803}{Phys. 
Rev. Lett. {\bf 105}, 146803 (2010)}.

\bibitem{Jiang2015} W. Jiang, Y-Y. Yang, A-B. Guo, {\it Study on magnetic properties of a nano-graphene bilayer}, \href{https://www.sciencedirect.com/science/article/pii/S0008622315301159}{Carbon, {\bf 95}, 190 (2015)}.

\bibitem{Jiang2016} W. Jiang, Y-N. Wang, A-B. Guo, Y-Y. Yang, K-L. Shi, {\it Magnetization plateaus and the susceptibilities of a nano-graphene sandwich-like structure}, \href{https://www.sciencedirect.com/science/article/pii/S0008622316307588}{Carbon, {\bf 110}, 41 (2016)}.

\bibitem{Wang2018} K. Wang, P. Yin, Y. Zhang, W.Jiang, {\it Phase diagram and magnetization of a graphene nanoisland structure}, \href{https://www.sciencedirect.com/science/article/pii/S0378437118303789}{Phys. A. {\bf 501}, 268 (2018)}.

\bibitem{Masrour2017_1} R. Masrour, A. Jabar, {\it Size effect in graphene nano-islands: A Monte Carlo study}, \href{https://link.springer.com/article/10.10072Fs10825-017-0990-y}{J. Comput. Electron, {\bf 16}, 576, (2016)}.



\bibitem{Masrour2017_2} R. Masrour, A. Jabar, {\it Magnetic properties of bilayer graphene: a Monte Carlo study}, \href{https://link.springer.com/article/10.1007/s10825-016-0930-2}{J. Comput. Electron, {\bf 16}, 12, (2017)}.



\bibitem{Jabar2016} A. Jabar, R. Masrour, {\it Magnetic properties of graphene structure: a Monte Carlo simulation}, \href{https://link.springer.com/article/10.1007/s10948-016-3417-2}{J. Supercond. Nov. Magn. {\bf 29}, 1363, (2016)}. 

\bibitem{Jabar2017} A. Jabar, R. Masrour, {\it Magnetic properties of a graphene with alternate layers}, \href{https://www.sciencedirect.com/science/article/pii/S0749603617320633}{Superlattices and Microstructures, {\bf 98}, 78, (2017)}.

\bibitem{Masrour2018} R. Masrour, A. Jabar, {\it Size and diluted magnetic properties of diamond shaped graphene quantum dots: Monte Carlo study}, \href{https://www.sciencedirect.com/science/article/pii/S0378437117313900}{Phys. A. {\bf 497}, 211, (2018)}.



\bibitem{Inglot2015}M. Inglot, V. K. Dugaev, and J. Barna\'{s}, {\it Thermoelectric and thermospin transport in a ballistic junction of graphene}, \href{http://journals.aps.org/prb/abstract/10.1103/PhysRevB.92.085418}{Phys. Rev. B {\bf 92}, 085418 (2015)}.

\bibitem{Dragoman2007} D. Dragoman, and M. Dragoman, {\it Giant thermoelectric effect in graphene}, \href{http://scitation.aip.org/content/aip/journal/apl/91/20/10.1063/1.2814080}{ Appl. Phys. Lett. {\bf 91}, 203116 (2007)}.

\bibitem{Zuev2009} Y. M. Zuev, W. Chang, and P. Kim, {\it Thermoelectric and magnetothermoelectric transport measurements of graphene},\href{http://journals.aps.org/prl/abstract/10.1103/PhysRevLett.102.096807}{Phys. Rev. Lett. {\bf 102}, 096807 (2009)}.

\bibitem{Wei2009} P. Wei, W. Bao, Y. Pu, C. N. Lau, and J. Shi, {\it Anomalous thermoelectric transport of Dirac particles in graphene},\href{http://journals.aps.org/prl/abstract/10.1103/PhysRevLett.102.166808}{Phys. Rev. Lett. {\bf 102}, 166808 (2009)}.

\bibitem{Kane1995}

\bibitem{Min2006} H. Min, J. E. Hill, N. A. Sinitsyn, B. R. Sahu, L. Kleinman, and A. H. MacDonald, {\it Intrinsic and Rashba spin-orbit interactions in graphene sheets}, \href{https://journals.aps.org/prb/abstract/10.1103/PhysRevB.74.165310}{Phys. Rev. B {\bf 74}, 165310 (2006)}.

\bibitem{Hernando2006} D. Huertas-Hernando, F. Guinea, and A. Brataas, {\it Spin-orbit coupling in curved graphene, fullerenes, nanotubes, and nanotube caps}, \href{https://journals.aps.org/prb/abstract/10.1103/PhysRevB.74.155426}{Phys. Rev. B {\bf 74}, 155426 (2006)}.

\bibitem{Yao2007} Y. Yao, F. Ye, X-L Qi, S-C Zhang, and Z. Fang, {\it Spin-orbit gap of graphene: First-principles calculations}, \href {https://journals.aps.org/prb/abstract/10.1103/PhysRevB.75.041401}{Phys. Rev. B {\bf 75}, 041401(R) (2007)}.

\bibitem{Yu2008} Yu. S. Dedkov, M. Fonin, U. R\"udiger, and C. Laubschat, {\it Rashba Effect in the Graphene/Ni(111) System}, \href{https://journals.aps.org/prl/abstract/10.1103/PhysRevLett.100.107602}{Phys. Rev. Lett. {\bf 100}, 107602 (2008)}.
 

\bibitem{Jin2006} D. Jin, Y. Ren, Z-Z. Li, M-W. Xiao, G. Jin, and A. Hu, {\it Spin-filter tunneling magnetoresistance in a magnetic tunnel junction} \href{http://journals.aps.org/prb/abstract/10.1103/PhysRevB.73.012414}{Phys. Rev. B {\bf 73} 012414 (2006)}. 

\bibitem{Beenakker2006} C.W.J. Beenakker, 
{\it Specular Andreev reflection in graphene}, \href{http://journals.aps.org/prl/abstract/10.1103/PhysRevLett.97.067007}{Phys. Rev. Lett. {\bf 97}, 067007 (2006)}.

\bibitem{Beenakker2008} C.W.J. Beenakker, {\it Colloquium: Andreev reflection and Klein tunneling in graphene}, \href{URL}{Rev. Mod. Phys. {\bf 80} 1337 (2008)}.

\bibitem{Liu2005} Y-X Li, Y. Guo, and B-Z. Li, {\it Rashba spin-orbit effect on electronic transport in ferromagnetic/semiconductor/ferromagnetic nanostructures under an applied electric field}, \href{https://journals.aps.org/prb/abstract/10.1103/PhysRevB.71.012406}{Phys. Rev. B {\bf 71} 012406 (2005)}.


\bibitem{Halterman2013} K. Halterman, O. Valls, and M. Alidoust, {\it Spin-Controlled Superconductivity and Tunable Triplet Correlations in Graphene Nanostructures}, \href{http://journals.aps.org/prl/abstract/10.1103/PhysRevLett.111.046602}{Phys. Rev. Lett. {\bf 111} 046602 (2013)}.


\bibitem{Beiranvand2016} R. Beiranvand, H. Hamzehpour, and M. Alidoust, {\it Tunable anomalous Andreev reflection and triplet pairings in spin-orbit-coupled graphene}, \href{https://journals.aps.org/prb/abstract/10.1103/PhysRevB.94.125415}{Phys. Rev. B {\bf 94}, 125415 (2016)}.

\bibitem{Beiranvand2017} R. Beiranvand, H. Hamzehpour, and M. Alidoust, {\it Nonlocal Andreev entanglements and triplet correlations in graphene with spin-orbit coupling}, \href{URL}{Phys. Rev. B {\bf 96}, 161403(R) (2017)}.


\bibitem{BeiranvandJAP2017} R. Beiranvand, H. Hamzehpour, {\it Spin-dependent thermoelectric effects in graphene-based superconductor junctions}, \href{URL}{J. App. Phys. {\bf 121}, 063903 (2017)}.


\bibitem{Jong1996} M.d. Jong, C.W.J. Beenakker, {\it Shot noise in mesoscopic systems}, \href{URL}{Arxiv, 9611140 (1996)}.

\bibitem{Tworzydlo2006} J. Tworzydło, B. Trauzettel, M. Titov, A. Rycerz, C.W.J. Beenakker, {\it Sub-Poissonian Shot Noise in Graphene}, \href{URL}{Phys. Rev. Lett. {\bf 96}, 246802 (2006)}.

\bibitem{DiCarlo2008} L. DiCarlo, J.R. Williams, Y. Zhang, D.T. McClure, C.M. Marcus, {\it Shot Noise in Graphene}, \href{URL}{Phys. Rev. Lett. {\bf 100}, 156801 (2008)}.

\bibitem{Danneau2008} R. Danneau, F. Wu, M.F. Craciun, S. Russo, M.Y. Tomi, J. Salmilehto, A.F. Morpurgo, P.J. Hakonen, {\it Shot Noise in Ballistic Graphene}, \href{URL}{Phys. Rev. Lett. {\bf 100}, 196802 (2008)}.



\end{thebibliography}
\end{document}